\documentclass[preprint,12pt]{elsarticle}

\usepackage{graphicx}

\usepackage{amssymb}

\usepackage{multirow}
\usepackage{tabu}
\usepackage{color}

\journal{International Journal of Plasticity}

\begin{document}
\newcommand{\etal}{{\it et al.}}

\begin{frontmatter}



\title{{\bf Temperature and high strain rate dependence of tensile deformation behavior in single crystal iron from dislocation dynamics simulations}}


\author{Meijie Tang and Jaime Marian}
\address{Lawrence Livermore National Laboratory}

\begin{abstract}
We conduct dislocation dynamics (DD) simulations of Fe periodic single crystals under tensile load at several high strain rates and temperatures. The simulations are enabled by the recent development of temperature-dependent dislocation mobility relations obtained from atomistic calculations. The plastic evolution in the simulations is governed by rapid initial dislocation multiplication, followed by a saturation of the flow stress when the subpopulation of slow plastic carriers becomes stabilized by dislocation annihilation. Above 500 K, edge dislocations coexist with screw dislocations and contribute proportionaly to the value of the flow stress. The DD simulations are used to interpret shock-loading experiments in Fe in terms of the relative importance of different strengthening mechanisms. We find that in the $10^4$-to-$10^6$ s$^{-1}$ strain rate regime, work hardening explains the hardening of shock-loaded bulk Fe crystals. 
\end{abstract}


\end{frontmatter}


\section{Introduction}
\label{intro}

The dependence of the flow stress with strain rate in many materials shows two clearly differentiated regimes \cite{i1,i2,i3,i4,i5,i6,i7,i8,i9,i10}, namely a slow increase up to strain rates on the order of $10^7$ s$^{-1}$ followed by a sharp upturn at higher rates. This shift is believed to be related to a transition in the mechanism of dislocation glide, namely from thermally-activated to viscous drag. Dislocation multiplication and glide may result in high levels of work hardening, often accompanied by twinning at the higher rates. This behavior has been confirmed experimentally for a wide number of materials with different underlying microstructures \cite{i2,i3,i7}. Multiscale integrated models that account for the relevant physical processes at each of the scales involved have proven very successful in predicting the general behavior of the material strength with strain rate \cite{barton2011,barton2012}. One such scale is that over which dislocation plasticity dominates the evolution of the strength. For studying this regime, methods capable of tracking millions of dislocation segments are of necessity. However, a stubborn bottleneck of existing models is the inclusion of the temperature dependence of different dislocation processes ---e.g.~glide and cross slip---  which are important to rationalize the 1D-to-3D deformation transition during shock loading of materials. 

Here, we perform dislocation dynamics (DD) simulations of homogeneous dislocation ensembles in body-centered cubic (bcc) Fe single crystals at different strain rates and temperatures to gain insight into the  mechanisms governing the flow stress in the strain rate regime thought to be dominated by dislocation glide.

A critical aspect for these simulations to be successful is the capability to capture the temperature and strain-rate dependence of single-dislocation motion. In particular, incorporating the dual character of screw dislocation mobility, with clearly distinct thermally-activated and viscous regimes, into DD simulations has remained a challenge for over two decades despite its undeniable importance. Several efforts aimed at addressing this shortcoming must be recognized \cite{tang1998,naa2012}, although only in a preliminary manner. Here we use dislocation mobilities derived from the calculations done by Gilbert \etal~\cite{gilbert2011} and Queyreau \etal~\cite{queyreau2011} using molecular dynamics (MD) simulations.

The paper is organized as follows. In the next section, we provide a brief description of the most salient features of DD, provide the simulation conditions, and describe the dislocation mobility functions in detail. We then provide results from the simulations, including stress-strain curves, dislocation density-strain curves and a comparison with available experimental data. We finalize with the discussion section and the conclusions. 

\section{Computational methods}
\label{comp}

All the simulations presented here were carried out using the Parallel Dislocation Simulator (ParaDiS) \cite{paradis}. The simulations comprised a cubic cell of size $L=5$ $\mu$m containing a fixed initial network dislocation density $\rho_0$.
To resemble a well-annealed crystal, in all cases we take $\rho_0=8.0\times10^{12}$ m$^{-2}$ consisting only of screw dislocations. Furthermore, all four $\small{\frac{1}{2}}\langle111\rangle$ Burgers vectors were populated equally in such a way as to produce zero net Burgers vector in the simulation cell. Network dislocations were made continuous and infinite in space by using periodic boundary conditions, and regenerative sources could only be created by interaction of the initial dislocation network with itself.

A simulation matrix was constructed with temperatures and strain rates of, respectively,  100, 300, and 600 K, and $10^4$, $10^5$ and $10^6$ s$^{-1}$. Tensile loading along the $[001]$ direction was performed in order to achieve multi slip conditions. Only glide on $\{110\}$ slip planes was considered. 
Fe is approximated to be a linear elastic solid and characterized by the values of the shear modulus $\mu=86$ GPa and Poisson's ratio $\nu=0.29$.

\subsection{Segment mobility}
The segment mobilities employed in this work are explained below. 

\emph{Screw mobility}. At very high strain rates, screw dislocation mobility is seen to be independent of its length \cite{dorn}. This can be rationalized in terms of the competition between kink pair nucleation vs.~propagation, which becomes comparable at these rates.
The relationship between stress $\tau$ and screw dislocation velocity $v$ in the thermally-activated and phonon drag regimes is:
\begin{eqnarray}
\tau_{th}(v;T)&=&\alpha\tau^{\ast}(T)\left\{\left[\frac{v}{C_0(T)}+v_0(T)\right]^\beta-v_0(T)^\beta\right\}\\
\tau_{ph}(v;T)&=&B(T)\left(v+v^{\ast}(T)\right)
\end{eqnarray}
where $T$ is the absolute temperature and the rest of parameters are given in Table \ref{tab:param}.
The contribution to the material strength from screw dislocations is obtained from the following rule of mixtures:
\begin{equation}
\tau_s=\left(\tau_{th}^n+\tau_{ph}^n\right)^{1/n}
\end{equation}

\begin{table}[h]
\caption{List of parameters and functional dependences for the screw dislocation mobility employed here.}
\centering
\begin{tabu} to 1.01\textwidth {|X[1.3,c] 
                                |X[7,l]<{\strut}
                                |X[2,c]<{\strut} 
                            |}\hline
{\small parameter} & {\small value} & {\small units} \\
\hline
$\alpha$ & 3.3 & - \\
$\beta(T)$ & $2.0\times10^{-4}T$ & - \\ 
$C_0(T)$ & $3710T^{-\frac{1}{2}}$ & m$\cdot$s$^{-1}$ \\
$v_0(T)$ & {\small $1.3$$\times$$10^{-10}T^3-1.5$$\times$$10^{-8}T^2-2.3$$\times$$10^{-6}T+2.5$$\times$$10^{-4}$} & - \\
$\tau^{\ast}(T)$ & $1200(1-0.001T)^2$ & MPa \\
$B(T)$ & {\small $-4.0$$\times$$10^{-2}T^3+38.3T^2-10^{4}T+1.7$$\times$$10^{6}$} & MPa$\cdot$s$\cdot$m$^{-1}$ \\
$v^{\ast}$(T) & {\small $4.5$$\times$$10^{-5}T^3-4.2$$\times$$10^{-2}T^2+10.2T-235.5$} & m$\cdot$s$^{-1}$ \\
$n$ & 10 & - \\
$A$ & 370.1 & {\small K$\cdot$m$\cdot$s$^{-1}$$\cdot$MPa$^{-1}$} \\
\hline
\end{tabu}
\label{tab:param}
\end{table}

\emph{Edge dislocations}. The edge dislocation mobility is assumed to follow a simple viscous law:
\begin{equation}
\tau_e=\frac{Tv}{A}
\end{equation}
where $A$ is also given in Table \ref{tab:param}.

\begin{figure}[h]
\centering
\includegraphics[width=1.0\linewidth]{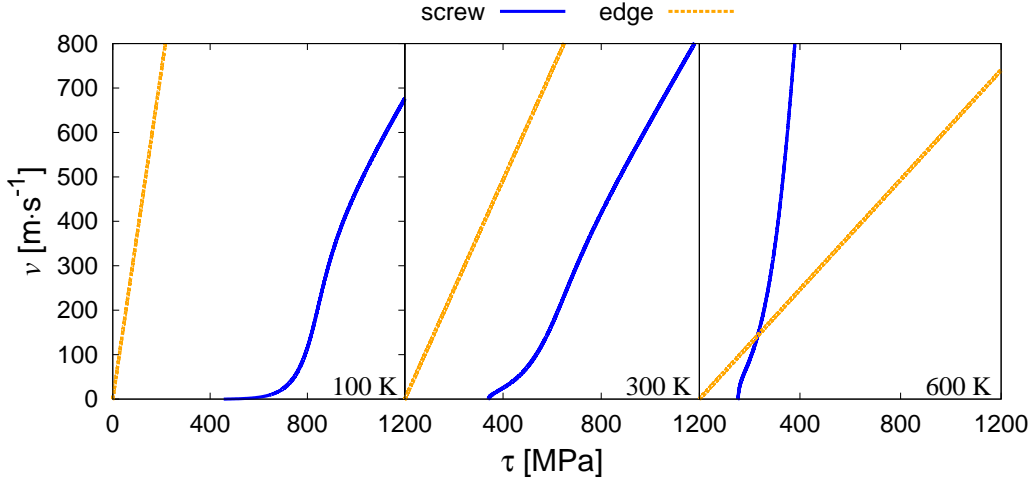}
\caption{Screw and edge dislocation mobilities at the three temperatures tested here.}
\label{mob}
\end{figure}

As a point of reference, with their mobilities described by the above functions, edge dislocations are three orders of magnitude faster than screw dislocations at 100 and two at 300 K, while at 600 K a crossover exists at 900 MPa, with edge velocities being slower than those of their screw counterparts. Figure \ref{mob} illustrates the behavior of the mobility function at the three temperatures of interest.

The above mobility function follows the so-called bcc0 construction \cite{bcc0a,cai2004}, which keeps edge dislocations confined to their slip plane $\vec{n}\equiv\vec{\xi}\times\vec{b}$ ($\vec{n}$: plane normal; $\vec{\xi}$: line tangent; $\vec{b}$: Burgers vector) while screw segments move on the plane dictated by the maximum resolved shear stress provided that such plane belong to the $\{110\}$ family. This makes screw dislocations free to cross-slip into any allowed slip plane. Climb is enabled, although with a mobility significantly smaller than that of edge dislocations, and it is not expected to play any significant role. 

\subsection{Numerical challenges}
The main numerical challenges are associated with the following aspects of the simulations:
\begin{itemize}
\item Time convergence of solution. We use an implicit Newton-Raphson method to integrate the equations of motion. Typically, convergence is achieved after a few iterations, with higher strain rates requiring more iterations.
\item High screw/edge mobility asymmetry. At low temperatures and/or strain rates, edge dislocation mobility may be up to several orders of magnitude higher than screw mobility. This sets restrictions on the time step duration, which may get as low as $10^{-11}$ to $10^{-13}$  s.
\end{itemize}

\section{Results}
\subsection{Numerical calculations}
\begin{figure}[h]
\centering
\includegraphics[width=1.0\linewidth]{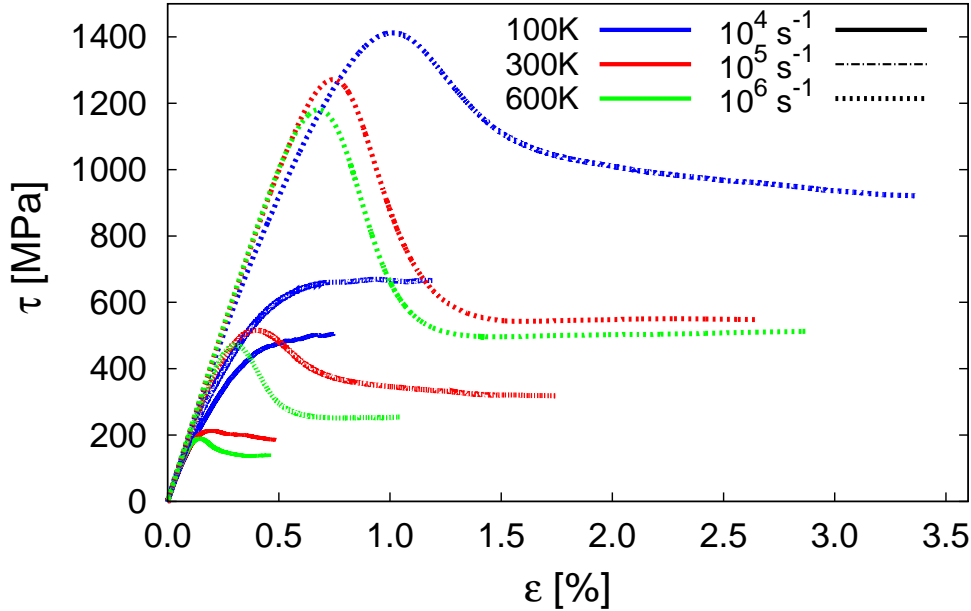}
\caption{Stress-strain curves for each of the nine simulation conditions considered. The simulations were carried out to the point of constant flow stress.}
\label{stress}
\end{figure}
Figure \ref{stress} shows the stress-strain curves for the combination of strain rates and temperatures considered in this work. The simulations were run up to the point of generally constant flow stress. This required varying amounts of computational time for each simulation, depending mostly on temperature.  By way of example, at 600 K, the simulations entailed 300,000 time steps and up to $1.5\times10^6$ CPU-hours.
In this work, the irreversible generation of heat by means of viscous plastic processes has been neglected. As a point of reference, Rittel \etal~\cite{ravi} have shown that in room temperature polycrystalline Fe specimens, the temperature increased only by a few degrees after 10\% deformation at $\dot{\varepsilon}\approx3.8\times10^3$ s$^{-1}$.

\begin{figure}[h]
\centering
\includegraphics[width=1.0\linewidth]{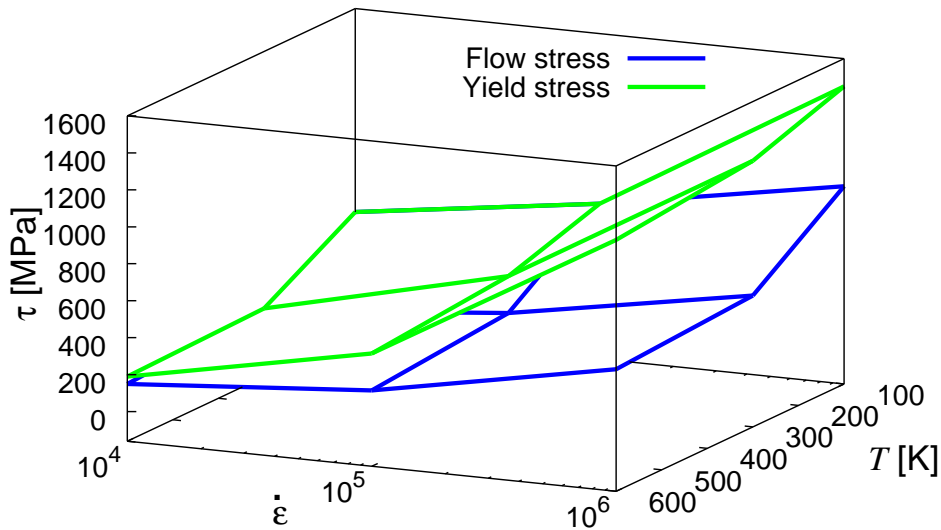}
\caption{Dependence of the yield and flow stress with temperature and strain rate.}
\label{3d}
\end{figure}

In some cases, we can distinguish a \emph{yield} stress, in the form of an initial hump, which is different from the steady state value of the flow stress. 
This corresponds to an elastic \emph{overshoot} in the stress response, as the initial dislocation density cannot accommodate the applied strain rate, leading to rapid dislocation multiplication at high rates that then lags the subsequent stress evolution.
The yield and flow stress surfaces are provided in Figure \ref{3d}. As the figure shows, this overshoot is facilitated by strain rate, temperature, or the compounded effect of both. 

\begin{figure}[h]
\centering
\includegraphics[width=1.0\linewidth]{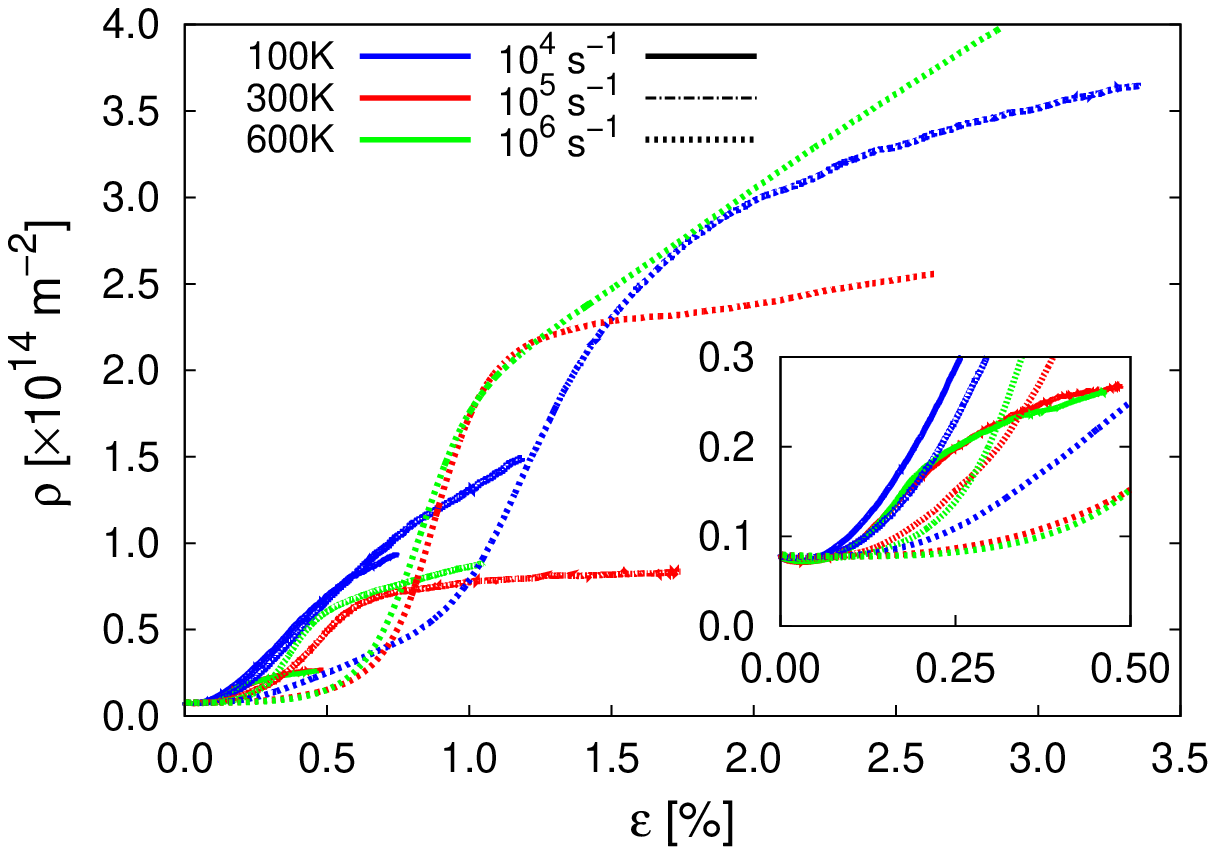}
\caption{Evolution of the total dislocation density with strain as a function of temperature and strain rate. The inset shows an amplification of the chart at low strains for visualizing the curves corresponding to $\dot{\varepsilon}=10^4$ s$^{-1}$.}
\label{rho}
\end{figure}

Figure \ref{rho} shows the evolution of the total dislocation density with strain. 
All the curves generally display three regimes: (i) a stationary one corresponding to the elastic regime; (ii) one of rapid growth following yield; and (iii) one of moderate multiplication rate when the flow stress has been established. The change is slope in going from (ii) to (iii) is more marked in the cases where the elastic overshoot is observed.

In general, plasticity is ultimately mediated by slow carriers of deformation. It is conventionally assumed that screw dislocations play that role, regardless of temperature. Under such an supposition, nonscrew segments and/or sources exit the simulation cell or exhaust themselves shortly after loading, resulting in a limited contribution to the overall plastic flow which is then controlled by screw dislocations. However, in this work we use dislocation mobilities that cover the entire temperature range, and, more importantly, that distinguish between the thermally-activated and viscous motion regimes for screw dislocations. As anticipated above, a crossover exists between the screw and edge dislocation mobilities. At 600 K, edge dislocations move slower than screws above 300 MPa, which is just below the transition stress that separates the thermally activated and viscous motion regimes.
Certainly, at $\dot{\varepsilon}=10^6$ s$^{-1}$ and 600 K, stress levels are always above 300 MPa and one then would expect plasticity to be equally partitioned among nonscrew and screw segments. Conversely, at 100 K, edge dislocations are unquestionably faster than screws at any stress, and so we then expect screw dislocations to bear the full weight of plastic flow.

To validate this line of thought, next we calculate the fraction of screw dislocations $f_s=\rho_s/(\rho_s+\rho_e)$ and track its evolution with strain. It can be seen that $f_s$ asymptotically converges to a rate-independent ---but temperature dependent--- value, which we term $f_{eq}$.
\begin{figure}[h]
\centering
\includegraphics[width=1.0\linewidth]{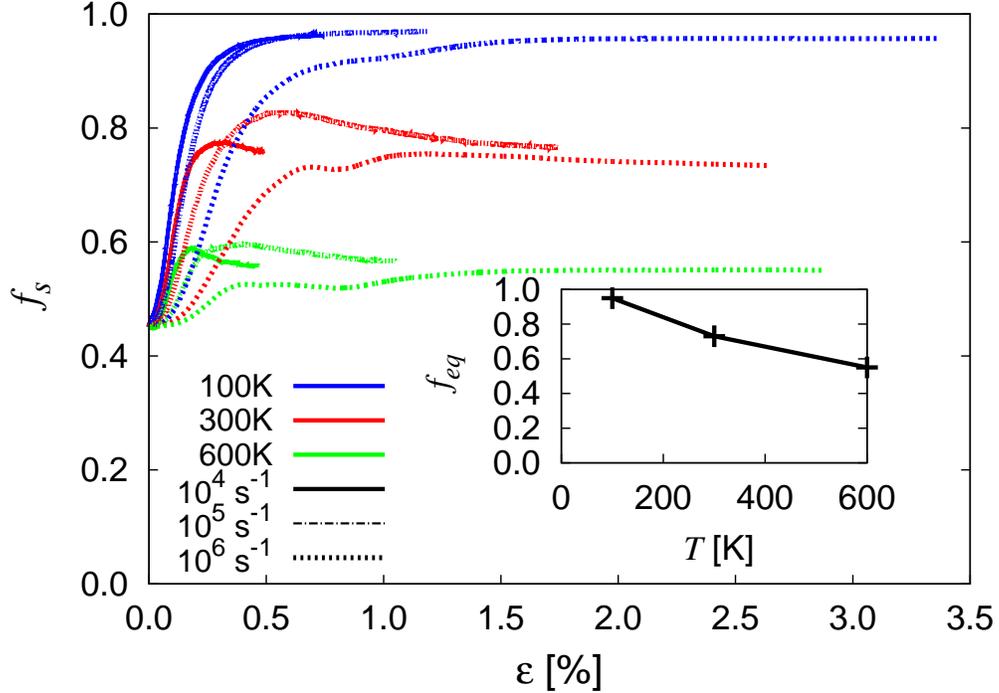}
\caption{Fraction of screw dislocations relative to the total dislocation density as a function of total strain. The inset shows the temperature variation of the asymptotic value $f_{eq}$.}
\label{lss}
\end{figure}
The respective values of $f_{eq}$ at 100, 300, and 600 K are 0.95, 0.73, and 0.55. Indeed, at 600 K there is almost an equipartition between edge and screw carriers, whereas at 100 K plasticity is clearly dominated by screw dislocation segments.

\subsection{Comparison with experiments}
On the basis of a relative large volume of experimental data, a number of researchers have rationalized the upturn in the flow stress for Fe at high strain rates as governed by the interplay between dislocation glide by a thermally activated mechanism versus glide by viscous drag \cite{follans1,follans2}. As well, in the present strain rate regime, twinning is known to take over dislocation glide as the dominant deformation mechanism at low temperatures in single crystal Fe \cite{meyers2001}. In this section, we compare the critical stresses obtained in our simulations with relevant experimental data to ascertain whether the mechanisms that control dislocation glide and interactions in DD capture the essential experimental behavior.
\begin{figure}[h]
\centering
\includegraphics[width=1.0\linewidth]{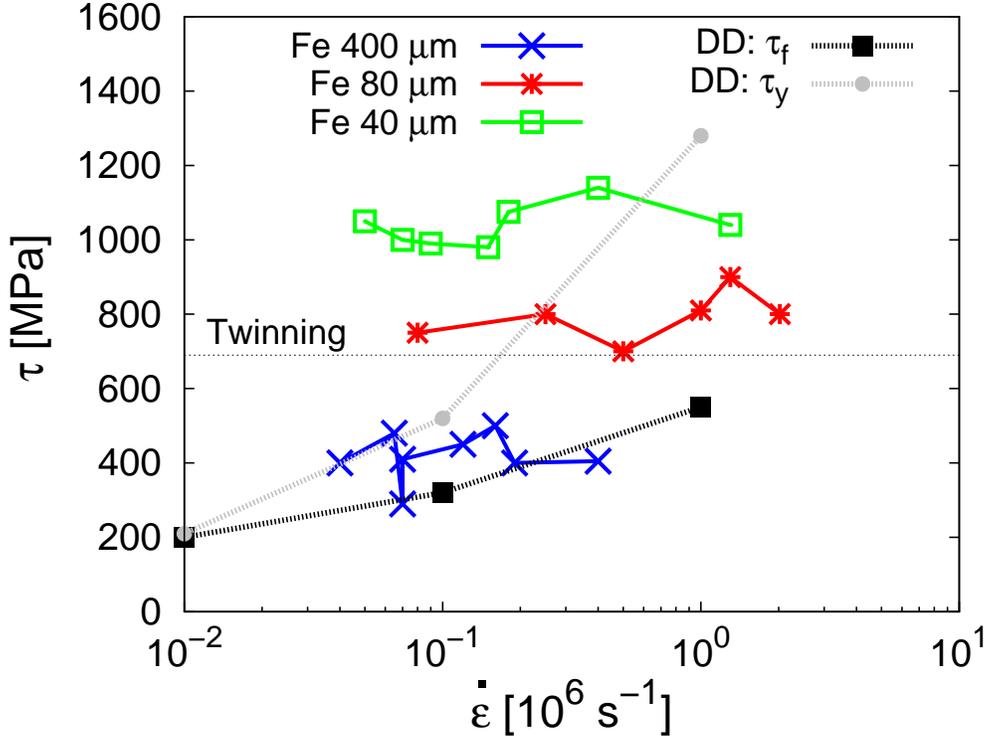}
\caption{Comparison of experimental data for ARMCO Fe with varying grain size \cite{arnold1994} and the yield and flow stresses from Fig.\ \ref{3d} at 300 K. The horizontal dashed line corresponds to the experimental twinning stress at room temperature \cite{ikeda1967}.}
\label{hel}
\end{figure}
Figure \ref{hel} shows the dependence of the Hugoniot elastic limit (HEL) of ARMCO Fe for three different grain sizes with strain rate \cite{arnold1994}. The HEL marks the limit after which irreversible (plastic) deformation is achieved in shock loading experiments, and, as such, reflects the response of the material to several possible deformation mechanisms. Conversely, DD simulations give the stresses derived from strain hardening (dislocation glide) processes. Evidently, our simulations correspond to bulk (single crystal) systems and, thus, the comparison with the data corresponding to a grain size of 400-$\mu$m is more meaningful. The comparison at room temperature is shown in Fig.\ \ref{hel}. The experimental twinning strength of single crystal Fe is also shown for reference \cite{ikeda1967}. Our results may provide evidence of glide-dominated conditions for large grain sizes, whereas twinning may be responsible for the strength in more fine-grained microstructures and, as mentioned, at lower temperatures as well \cite{meyers2001}.

\section{Discussion}

\subsection{Constitutive modeling}


Models of materials strength under high strain rates with various degrees of phenomenology have been proposed (see reviews \cite{voy2002,armstrong2008,gray2012}. Typically, these models focus on one or more aspects of dislocation glide from a micro mechanical point of view\cite{follans1,follans2,za}. However, although these approaches have been successful in explaining semiquantitatively many aspects of dynamic deformation in many materials, they lack physical connection across all the temporal and spatial scales involved.

Recently, Barton \etal~have provided a self-consistent multiscale model of strength in bcc metals under fast loading rates that takes into account both thermally-activated and drag contributions to plastic flow \cite{barton2011,barton2012}. Within the multiscale model, the dislocation density evolution is assumed to arise from expansion of existing dislocation segments and dislocation annihilation, and it that sense it is consistent with the mechanisms included in the underlying dislocation dynamics calculations:
\begin{eqnarray}
\rho_{\rm sat}&=&\rho_0\left(\frac{\dot\varepsilon}{\dot\varepsilon_N}+S_0\right)\\
\dot\rho&=&R\left(1-\frac{\rho}{\rho_{\rm sat}(\dot\varepsilon)}\right)\dot\varepsilon
\end{eqnarray}
where $\dot\varepsilon$ is the prescribed strain rate, $\rho$ is the total dislocation density, $\rho_{\rm sat}$ is the strain-rate dependent saturation density, and $\rho_0$, $\dot\varepsilon_N$, $S_0$ and $R$ are fitting parameters determined via DD simulations\footnote{We have not attempted to compute these here.}.
In the context of these multiscale models, our simulations are a critical piece that connects materials strength to the underlying physical dislocation processes obtained at the microscopic scales.

\subsection{Physical implications}

As discussed earlier, our simulations capture only the work hardening component of the total material strengthening, which may consist of thermal activation and phonon drag contributions. As such, they are typically integrated into multiscale models as the one described in the previous subsection. However, in the context of single crystal strength, our simulations alone are sufficient to explain some aspects of the dependence of the flow stress with strain rate. Our results for the flow stress at 300 K  plotted in Fig.\ \ref{hel} are in good agreement with the experimental results for 400-$\mu$m polycrystalline Fe at room temperature by Arnold \cite{arnold1994}, which can be considered as the most representative of bulk behavior. Our data is also in general agreement with the measurements by Weston \cite{weston1992} for REMCO Fe. However, ref.\ \cite{weston1992} does not include any microstructural information of the as-received material and thus, for the sake of caution, we do not include the data in Fig.\ \ref{hel}.
This agreement may be an indication that work hardening is the predominant strength contribution in bulk Fe crystals at rates at or below $10^6$ s$^{-1}$.
For smaller grain sizes, or at higher simulation temperatures, the material strength surpasses the twinning strength of $\approx$700 MPa \cite{ikeda1967,meyers2001}, which blurs the strain rate sensitivity picture by enabling other deformation mechanisms. Our simulations at 300 K also reveal a threshold stress around 200 MPa as the thermally activated glide limit.

The elastic overshoot in the stress response observed at some combinations of temperature and strain rate is caused by the initial dislocation density being incapable of coping with the applied strain rate. In this sense, including (homogeneous) dislocation nucleation may help ease the burdens of strain relief on the original dislocation density imposed by the boundary conditions \citep{sheha2006,sheha2012,gurru}. However, this overshoot has been routinely seen experimentally as well \cite{ravi,weston1992}, so that the role of dislocation nucleation may not be as determining as initially thought.

Finally, we touch on the issue of the dislocation density partition between fast and slow carriers. The results in Fig.\ \ref{lss} show that the relative fraction of screw dislocations is solely a function of temperature, as dictated by the respective mobilities (cf.\ Fig.\ \ref{mob}), and independent of the strain rate. The data are conclusive in showing that, below room temperature, the population of slow carriers --responsible for plasticity-- is dominated by screw dislocation segments. Under these circumstances, thermally activated plastic flow cannot be neglected, which is consistent with the upturn traditionally seen in the $\sigma$-$\dot\varepsilon$ relation at very high strain rates. At higher temperatures, phonon drag is sufficiently important to slow down non-screw segments and all dislocations contribute noticeably to plasticity.

\subsection{Limitations of the simulations}
Next, we discuss the potential shortcomings of our approach. Evidently, treating Fe as an isotropic elastic material is the first limitation. However, it is one by construction, and widely seen as acceptable for discrete dislocation dynamics simulations. There are also two limitations associated with the high speed deformation conditions: (i) the use of static field solutions when the dislocation velocities are on the order of shear wave velocity, and (ii) neglecting inertial effects. These effects have also been discussed in the literature (e.g.~refs.\ \cite{gurru} and \cite{follans2}, respectively) and are thought to have only a limited impact when the dynamics are governed by dislocation multiplication and annihilation, as is the case here.
Finally, we have already mentioned the inclusion of homogeneous nucleation sources to facilitate dislocation density increases commensurate with the rapid stress increases. However, in the regimes studied here, at relatively low stresses, the impact of this contribution is expected to be only marginal.

\section{Summary}
We have carried out a systematic study of uniaxial high-rate loading in single crystal Fe as a function of temperature and strain rate. These simulations have been enabled by recent temperature-dependent dislocation mobility functions obtained from atomistic calculations. Our main conclusions are:
\begin{itemize}
\item Our simulations show that strengthening scales directly with strain rate and inversely with temperature. Under certain conditions, an elastic overshoot is observed, caused by the inability of existing dislocation sources to cope with the imparted strain rate. The existence of this overshoot in experiments in Fe may be an indirect indication that dislocation nucleation does not play a very important role, which further substantiates our results. 
\item The total dislocation density initially grows quadratically with strain --an indication of the dominance of dislocation multiplication-- and then stabilizes into a steady state saturation value.
\item The evolution of the ratio of screw dislocations relative to the total dislocation density is seen to be independent of strain rate and inversely proportional to temperature. This ratio is indicative of the importance of nonscrew carriers at each temperature and strain rate.
\item In the strain rate regime explored here, good agreement between experiments and simulations may be indicative that strain hardening is the dominant strengthening mechanism in Fe in the strain rate regime explored.
\end{itemize}

\section*{Acknowledgments}
his work was performed under the auspices of the U.S. Department of Energy by Lawrence Livermore National Laboratory under Contract No.\ DE-AC52-07NA27344. J. M. acknowledges support from DOE's Early Career Research Program.




\bibliographystyle{model1-num-names}
\bibliography{<your-bib-database>}



\end{document}